\documentclass[final,5p,times,twocolumn]{elsarticle}
\def\fw{\linewidth}

\usepackage{amssymb}
\usepackage{amsmath}
\usepackage{natbib}
\usepackage{algorithmic}
\usepackage{algorithm}

\def\nbody{\ {\nbod-body}\ }

\newcommand{\nbs}{\mbox{$n_{\rm blocksize}$}}
\newcommand{\nspb}{\mbox{$n_{\rm systems/block}$}}
\newcommand{\ncs}{\mbox{$n_{\rm chunksize}$}}
\newcommand{\nws}{\mbox{$n_{\rm warpsize}$}}
\newcommand{\nthread}{\mbox{$n_{\rm thread}$}}

\newcommand{\nsys}{\mbox{$n_{\rm sys}$}}
\newcommand{\npl}{\mbox{$n_{\rm pl}$}}
\newcommand{\nbod}{\mbox{$n$}}

\newcommand{\go}{\ {\raise-.5ex\hbox{$\buildrel>\over\sim$}}\ }
\newcommand{\lo}{\ {\raise-.5ex\hbox{$\buildrel<\over\sim$}}\ }
\newcommand{\be}{\begin{equation}}
\newcommand{\ee}{\end{equation}}
\newcommand{\bea}{\begin{eqnarray}}
\newcommand{\eea}{\end{eqnarray}}

\newcommand{\jorg}[1]{#1}

\def\nv{NVIDIA}
\def\spt{{\it system per thread}}
\def\Spt{{\it System per thread}}
\def\bpt{{\it body per thread}}
\def\Bpt{{\it Body per thread}}
\def\cpt{{\it component per thread}}
\def\Cpt{{\it Component per thread}}
\def\bppt{{\it body-pair per thread}}
\def\Bppt{{\it Body-pair per thread}}

\begin{document}

\begin{frontmatter}

\title{Swarm-NG: a CUDA Library for Parallel \nbody Integrations
with focus on Simulations of Planetary Systems\tnoteref{source}}

\author[CISE]{Saleh Dindar\corref{cor1}}
\ead{saleh@cise.ufl.edu}

\author[ASTRO]{Eric B. Ford\corref{cor2}}
\ead{eford@astro.ufl.edu}
\author[LSST,STEW]{Mario Juric}
\author[CISE]{Young In Yeo}
\author[CISE]{Jianwei Gao}
\author[ASTRO]{Aaron C. Boley}
\author[ASTRO]{Benjamin Nelson}
\author[CISE]{Jorg Peters}

\cortext[cor1]{Primary corresponding author}
\cortext[cor2]{Secondary corresponding author}

\address[CISE]{Department of Computer \& Information Science \& Engineering, University of Florida, CSE Building, Gainesville, FL, 32611-6120, USA}
\address[ASTRO]{Department of Astronomy, University of Florida, 211 Bryant Space Science Center, PO Box 112055, Gainesville, FL, 32611-2055, USA}
\address[LSST]{LSST Corporation, 933 N. Cherry Avenue, Tucson, AZ 85721, USA}
\address[STEW]{Steward Observatory, University of Arizona, 933 N. Cherry Avenue, Tucson, AZ 85721, USA}

\tnotetext[source]{Source code available at {http://www.astro.ufl.edu/{$\scriptstyle\mathtt{\sim}$}eford/code/swarm/}}

\begin{abstract}

We present Swarm-NG, a C++ library 
for the efficient direct integration of many \nbody systems 
using a Graphics Processing Unit (GPU),
such as \nv's Tesla T10 and M2070 GPUs.
While previous studies have demonstrated the benefit of GPUs for \nbody
simulations with thousands to millions of bodies, 
Swarm-NG focuses on \emph{many few-body} systems, e.g., thousands
of systems with 3\ldots 15 bodies each, as  is typical for the study of
planetary systems.
Swarm-NG parallelizes the simulation, including both the numerical 
integration of the equations of motion and the evaluation of forces
using \nv's ``Compute Unified Device Architecture'' (CUDA) on the GPU.
\jorg{
Swarm-NG includes optimized implementations of 4th order time-symmetrized
Hermite integration and mixed variable symplectic integration,
as well as several sample codes for other algorithms to illustrate
how non-CUDA-savvy users may themselves introduce customized integrators
into the Swarm-NG framework.}
%
To optimize performance, 
we analyze the effect of GPU-specific parameters on performance
under double precision.
%
For an ensemble of  131072 planetary systems,
each containing 3 bodies, the \nv\ Tesla M2070 GPU
outperforms a 6-core Intel Xeon X5675 CPU
by a factor of $\sim 2.75$. 
Thus, we conclude that modern GPUs offer an attractive alternative
to a cluster of CPUs for the integration of an ensemble of many
few-body systems.
 
Applications of Swarm-NG include studying the late stages of planet formation, 
testing the stability of planetary systems and evaluating the goodness-of-fit
between many planetary system models and observations of extrasolar planet 
host stars (e.g., radial velocity, astrometry, transit timing).  
While Swarm-NG focuses on the parallel integration of many planetary systems, 
the underlying integrators could be applied to a wide variety of problems that
require repeatedly integrating a set of ordinary differential equations
many times using different initial conditions and/or parameter values.

\end{abstract}

\end{frontmatter}

\section{Introduction}

\subsection{Background}
An \nbody simulation numerically approximates the evolution of a system of
bodies in which each body continuously interacts with every other body,
a fundamental component of many physical and chemical systems.
%
%
\nbody simulations are ubiquitous in astrophysics and planetary science.  Example applications include investigating the trajectories of spacecraft, the formation and orbital evolution of the solar system and other planetary systems, the delivery of water to Earth via collisions with asteroids and/or comets, the evolution of star clusters, the formation of galaxies and even the evolution of the entire universe.  

Given the widespread use of \nbody simulations, astronomers have developed a variety of algorithms and computer programs for performing \nbody integrations.  
At its core, an \nbody simulation requires solving a set of $3\times\nbod$~ second-order ordinary differential equations (ODEs).  
For $\nbod\ge3$, most sets of initial conditions result in chaotic evolution and are best studied numerically.
The computational requirements of \nbody simulations can be significant, either due to long timescales 
(e.g., billions of years), a large number of bodies (e.g., $\sim10^{5\ldots 7}$ for star cluster,
$>10^{9}$ for galaxy or universe), and/or the need to consider a large number of systems with slightly
different initial conditions (e.g., $\sim10^{6\ldots 9}$ model evaluations for Bayesian analysis of exoplanet observations). 

%
%
Several previous studies have demonstrated that the combination 
of modern Graphical Processing Unit (GPU) hardware and the CUDA (Compute Unified Device Architecture) programming environment can greatly accelerate a gravitational \nbody problem for large $\nbod$ (e.g., star clusters)
\cite{
2012JCoPh.231.2825B, 
Belleman2008,
Capuzzo-Dolcetta2011,
Gaburov2009,
Konstantinidis2010,
Hubert:2007:Gem3,
Nitadori2012,
Zwart2007}.
%
%
Here we focus on the problem of integrating an ensemble of \emph{many few-body} systems,
e.g., thousands to millions of systems with 3 to tens of bodies each, as is needed for the study of planetary systems.  

Given the large number of integrations of few-body
systems, GPUs can dramatically reduce the total
 time required to obtain scientific results for many real-world
applications (\S\ref{Sec:Applications}).








In this paper, we present the Swarm-NG library for parallel integration of \nbody systems.  
In \S\ref{Sec:Problem}, we describe the physical setup and 
the numerical methods. 
In \S\ref{Sec:Cuda}, we describe key aspects of GPU computing with CUDA.
In \S\ref{Sec:Implementation}, we discuss the details of our implementations.
In \S\ref{Sec:Results}, we present performance benchmarks. 
In \S\ref{Sec:Discussion}, we discuss present and likely applications.

\section{Physics and Numerics of \nbody systems}\label{Sec:Problem}

\subsection{Physics}
\label{Sec:Physics}
The time evolution of a classical system of \nbod\
bodies is described by Newton's laws of motion,
${\bf F_i} = m_{\bf i}{\bf a_i} $, 
where ${\bf a_i}$ is the acceleration vector (second time derivative of the position vector, ${\bf x_i}$)
for the $i$th body, ${\bf F_i}$ is the gravity force of other bodies
on the $i$th body 
and $m_{\bf i}$ is the mass of the $i$th body.
Swarm-NG integrates systems interacting under Newtonian gravity, 
\begin{equation}
   {\bf a_i} =
   \sum_{j\ne~i} \frac{G m_j ({\bf x_j} - {\bf x_i})}
   { \left| {\bf x_j} - {\bf x_i} \right|^{3} },
\end{equation}
where $m_j$ is the mass of the $j$th body and 
$G$ is Newton's Gravitational constant.  
Without loss of generality, we set $G=1$. 
 If we use the astronomical unit (AU) as the unit of length and
 the solar mass ($M_\odot$) as the unit of mass, then one year 
corresponds to $2\pi$ time units.  
The modular design of Swarm-NG allows for the Newtonian force law
 to be replaced with an alternative force law, 
should users want to include additional effects such as general relativity,
 precession due to oblateness, or gas drag.

\subsection{Numerical Integration of ODEs}
\label{Sec:ODEs}
For numerical integration, we replace the system of 3$\times$\nbod\
second-order differential equations by 6$\times$\nbod\
first-order differential equations,
\begin{align}
   \notag
   {\bf \dot{v_i}} & = {\bf a_i} \\
   {\bf \dot{x_i}} & = {\bf v_i}, 
   \label{eq:newton}
\end{align}
where ${\bf v_i}$ is the velocity of the $i$th body.

There are a variety of methods for numerically integrating 
\eqref{eq:newton}.
%
The simplest, Euler integration given by
\begin{align*}
	{\bf {v_i}}(t+\Delta t)  &:= {\bf {v_i}}(t) + {\bf a_i}(t) \Delta t,\\
	{\bf {x_i}}(t+\Delta t)  &:= {\bf {x_i}}(t) + {\bf v_i}(t) \Delta t + \frac{1}{2} {\bf a_i}(t) {\Delta t}^2,
\end{align*}
is however famously unstable and not practical for scientific simulations.  
Swarm-NG provides several advanced integration algorithms.  
The Hermite (\S\ref{Sec:HermiteTheory}) and Mixed Variables
Symplectic (MVS; \S\ref{Sec:MvsTheory}) integrators are currently
the workhorse for scientific short-term and long-term integrations, 
respectively.  
Other integrators (e.g., Verlet, Runge-Kutta, modified midpoint method) 
have been implemented in Swarm-NG to compare their efficiency when
using GPUs and to ensure that the Swarm-NG framework is sufficiently 
general to accommodate a variety of complex integration algorithms.  

The Swarm-NG libraries are designed so that advanced users can quickly
implement a new integration algorithm by writing code to advance the
system by one time-step (i.e., a ``propagator'' class).  They can then
decide whether to write a more efficient but less generic and more
complicated ``integrator'' class, where user has more control over the
overall algorithm.  Converting the Hermite propagator into a Hermite
integrator class resulted in more complex code and four additional
lines of code, but provided a performance gain of up to 5\%, depending
on ensemble parameters.

While our implementation of these algorithms on the GPU is original, the 
underlying integration algorithms are the same as the corresponding 
standard CPU-based integrators. 
Therefore, readers may consult standard texts and references
on the accuracy and stability of each algorithm to determine 
which algorithms are appropriate for their scientific problem.

\subsubsection{Hermite}
\label{Sec:HermiteTheory}
The fourth-order, time-symmetric Hermite integrator is described
in \cite{Kokubo1998}.  
Like a standard Hermite integrator, it uses analytic expressions to calculate both the acceleration and ${\bf j} := \dot{\bf a}$, the ``jerk'' (time derivative of the acceleration).  Each Hermite step consists of a prediction step,
\begin{align}
\notag
{\bf x_1} & :=  {\bf x_0} + {\bf v_0} \Delta t + \frac{1}{2} {\bf a_0} (\Delta t)^2 + \frac{1}{6} {\bf j_0} (\Delta t)^3 \\
{\bf v_1} & :=  {\bf v_0} + {\bf a_0} \Delta t + \frac{1}{2} {\bf j_0} (\Delta t)^2, 
\end{align}
followed by one or more correction steps,
\begin{align}
\notag
{\bf x_{k+1}} & :=  {\bf x_k} - \frac{3}{10} \left( {\bf a_0}-{\bf a_k} \right) (\Delta t)^2 - \frac{1}{60} \left( 7 {\bf j_0} + 2 {\bf j_k} \right) (\Delta t)^3\\
{\bf v_{k+1}} & :=  {\bf v_0} - \frac{1}{2} \left( {\bf a_0} - {\bf a_k} \right) \Delta t - \frac{1}{12} \left( 5 {\bf j_0} +  {\bf j_k} \right) (\Delta t)^2,
\end{align}
where the ${\bf a_k}$ and ${\bf j_k}$ are recalculated based on ${\bf x_k}$.
Following \cite{Kokubo1998}, Swarm-NG's Hermite integrator stops after two 
iterations of the correction step.

We implement two versions of the Hermite integrator, one with a fixed time-step and one with an adaptive time-step (see \S\ref{Sec:HermiteImpl}).
Swarm-NG also includes a CPU-based Hermite integrator (fixed time step only), allowing for direct comparisons of CPU and GPU performance for this algorithm.

\subsubsection{Runge-Kutta}
\label{Sec:RKTheory}
Swarm-NG provides a fourth/fifth-order Runge-Kutta Cash-Karp integrator closely
based on the \texttt{gsl\_odeiv2\_step\_rkck} stepper from the
GNU Science Library.   
We implemented two versions of the Runge-Kutta integrator, one with a 
fixed time-step and one with an adaptive time-step (see \S\ref{Sec:RKCKImpl}).  
The Runge-Kutta integrator uses only the accelerations and not the jerk.  
In order to achieve fifth order accuracy, 
the integrator estimates higher derivatives numerically.  
Therefore, both versions of the Runge-Kutta integrator require six sub-steps 
and six force evaluations for each step (fixed time-step) or trial step 
(adaptive time-step).  
Thus, this integrator is significantly more computationally expensive 
and memory intensive than the Hermite integrator.  
Its primary advantage is that only the acceleration needs to be specified. 
This makes it is easier to use with alternative force laws,
when deriving an analytic expression for the jerk is not practical.  

\subsubsection{Mixed-Variable Symplectic}\label{Sec:MvsTheory}
The Mixed-Variable Symplectic (MVS) integrator is optimized for integrating
systems containing one dominant central body (e.g., star) and several small 
mass bodies (e.g., planets). 
Technically, it is not based on integrating the exact equations of motion, 
but rather the evolution of a map which has very similar properties to the 
actual equations of motion.  
The theory of the MVS integrator \cite{Wisdom1991} is beyond the scope
of this paper.
We provide only an overview of the calculations and the essential properties
so that readers can begin to understand the parallelization and judge
whether MVS is an appropriate integrator for their problem.

Each step of the MVS integrator requires three kinds of sub-steps.
To minimize the error due to the symplectic nature of the algorithm, the 
sub-steps are broken in half steps and calculated symmetrically as:
\begin{itemize}
	\item{Cartesian drift half step}
	\item{Kick half step}
	\item{Keplerian drift step}
	\item{Kick half step}
	\item{Cartesian drift half step}
\end{itemize}
In the \emph{Cartesian drift half steps}, the position of the center of mass is advanced based
on the current center-of-mass velocity and the position of each planet
relative to the central body is advanced due to the reflex motion of
the central body in response to all of the planets.\\
In the \emph{Interaction or Kick half steps}, the velocities of each planet are
advanced based on the acceleration due to the mutual gravitational interaction
of the planets (while excluding the acceleration due to gravity between each 
planet and the central body).  \\
In the \emph{Keplerian drift step}, the positions and velocities of each planet
are evolved along a Keplerian orbit (i.e., elliptical path with a
non-uniform speed given by Kepler's third law of planetary motion)
corresponding to the trajectory of the planet if it were not being
perturbed by any other planets.  

As the name suggests, the MVS integrator makes use of multiple 
parameterizations of the system state.
There is a one-to-one mapping between Cartesian coordinates 
and Keplerian orbital parameters (e.g., \cite{Murray2000}).  
Using the two sets of variables makes each sub-step extremely simple. 
For example, the $i$th planet can be advanced exactly along a Keplerian
orbit for an arbitrary $\Delta t$ simply by increasing the mean anomaly 
by $\frac{2\pi\Delta t}{P_i}$, where $P=\left(\frac{G(m_0+m_i)}{a_i^3}\right)^{\frac{1}{2}}$ is the orbital 
period, $a_i$ is the semi-major axis, $m_0$ is the mass of the central body 
and $m_i$ is the planet mass.  
The work of the MVS integrator is dominated by converting between the two
representations of the system.
For few-body systems, the coordinate conversions can even require
more computation than the force calculation.  
The most expensive part of the conversion is iteratively solving 
Kepler's equation that relates the orbit in space to the orbit in time.
Swarm-NG adopts a previously developed CUDA kernel for repeatedly solving
Kepler's equation on the GPU \cite{Ford2009}.

While the MVS integrator is only second-order in the time step, 
the error term for the integration is also proportional to the ratio
of the planet masses to the central mass.  
The Keplerian drift sub-step evolves planets along Keplerian orbits, allowing accurate integrations 
even for large time-steps.  
If there were no mutual planetary perturbations 
then one could use an arbitrarily long time-step and still maintain accuracy
to machine precision.
Even more importantly, the MVS integrator is symplectic,
resulting in excellent long-term energy conservation and only linear growth 
in the error of the orbital phase.  
This makes the MVS integrator very well suited for long-term integrations
of planetary systems (e.g., testing for long-term stability).  

The main drawback of MVS is that, to maintain the symplectic nature,
a fixed time-step is required.
This makes the MVS integrator inappropriate for evolving systems 
through close encounters, which would require either an adaptive time-step
scheme or an unrealistically small fixed time-step.

\section{The \nv\ CUDA Architecture }
\label{Sec:Cuda}

A GPU offers between 5 and 30 \emph{multiprocessors} 
with Single Instruction, Multiple Thread (SIMT) architecture.
A \emph{kernel} is a portion of an application program, compiled to the 
instruction set of the GPU and loaded to the GPU for execution on one 
or more multiprocessors.
To achieve high parallelism, the kernel is launched on a large number of 
compute \emph{threads}, each operating on its own set of data. 
Threads are organized into a grid of \emph{thread blocks}, each containing 
$\nbs$ threads. 
CUDA-capable GPUs
have an on-chip \emph{shared memory} with very fast general 
read and write access, approximately as fast as on-chip registers.  
This memory is shared among threads within the same block. 
%
The local and global memory space are implemented as read-write 
regions of \emph{device memory}
which has high throughput, but also high latency.
While this memory is not cached on earlier generations of GPUs (before Fermi),
recent GPU architectures (Fermi and Kepler) include a two level cache mechanism;
see Table \ref{Tab:GpuSpecs}.
Still, the high computational throughput means
that memory access can easily become the bottleneck.  




\subsection{Challenges of GPU Computing}
GPU memory is arranged in different memory spaces, e.g., 
the latency of accessing global memory is about 100
times that of accessing shared memory or registers. 
In order to reduce the memory latency, we need to maximize the use of the
available bandwidth for memory with lower latency. Therefore a well-designed memory
access pattern can be the key to maximizing the instruction throughput.
Prior to Fermi-based GPUs, many CUDA kernels minimized memory access cost
by loading data from device memory into shared memory before
processing. 
Using the L1 cache on Fermi-based chips, many kernels now achieve a
similar performance without using the shared memory as a user-managed cache. 
Shared memory is nevertheless valuable when sharing data between threads.
Given the high computational throughput, GPU-based codes are often memory bandwidth bound,
limiting performance by the number of registers per thread, 
shared memory per multiprocessor and/or size of the L1 cache on a given GPU
(see Table \ref{Tab:GpuSpecs}).

Transfer between the device memory and the host memory is even slower
than between the multiprocessors and device memory. Therefore, Host-Device
communications should be minimized. This brings us the trade off between data
control and parallelism. That is, for a better utilization of device, we
need to move more code from the host to device, even if that means sometimes running
kernels that included small sections of serialized computations.





\section{Implementation}
\label{Sec:Implementation}

In this Section, we describe implementations of Swarm-NG integrators.
First, we 
describe how the computation is distributed among CUDA cores
in section \S\ref{Sec:WorkModel}. 
%
Next, implementation of gravitational force calculation is discussed in \S\ref{Sec:Forces}.
After discussing the concepts, in \S\ref{Sec:Integrators}, we discuss specific 
details about implementation of different numerical methods
which covers integrator algorithms and calculation of time step
and gravitational force.
Finally, in \S\ref{Sec:Optimisations}, we 
describe how the computation is distributed among CUDA cores
and techniques we use to minimize the effects of memory latency.
we also enumerate
several techniques we used to optimize performance for CUDA Fermi 
architecture.

\subsection{Work Parallelization Models}\label{Sec:WorkModel}
\begin{figure}[htb]
     \includegraphics[width=\fw]{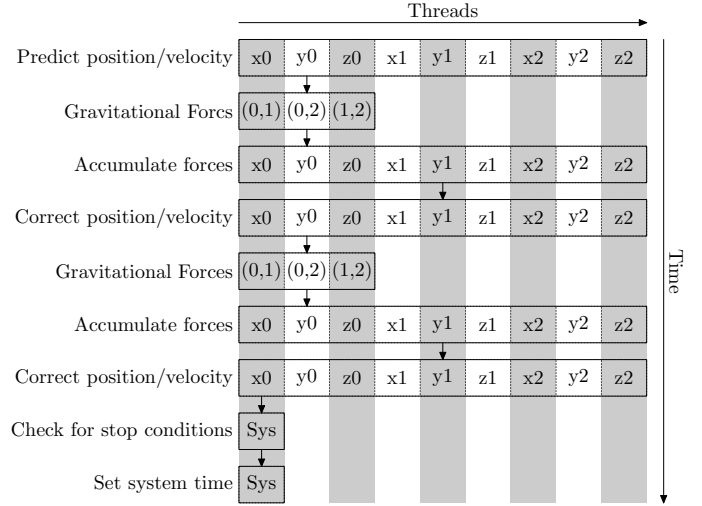}
     \caption{Illustration of effort distribution across threads with a multiprocessor for a single system with two planets using the Hermite integrator and default gravitation class.  The notation x1..z2 indicates the component and body number that is being assigned as a part of current task.  (i,j) pairs indicate that the interaction between bodies i and j being assigned to a thread as a part of current task.  ``Sys'' refers to assignment of the current task to one thread to operate on the whole system.
     The execution changes parallelism for each substep. The columns represent the threads that are running and rows represent the tasks being scheduled. The active threads are shown in the boxes.  For some steps, not all threads get assigned a task and some sit idle.  Active threads from different systems are grouped into a single warp, allowing the GPU scheduler to minimize the number of idle threads.   }
	\label{fig:Flow}
\end{figure}

The raw computational power of modern GPUs can result in performance being limited by  memory transfer.
Therefore, a general rule of thumb for optimizing GPU code is to implement parallel algorithms that hide memory latency by providing the GPU enough work to remain well-utilized for one set of calculations while waiting for data for a subsequent set of calculations. 

Our original implementation of Swarm-NG (v0.1) used only one thread per system.  
This work model led to a simple parallelization and was quite efficient for large ensembles of small systems using the  Hermite integrator.
However, fully utilizing the GPU required a large ensemble of systems and that each thread use a significant amount of memory, even for the Hermite integrator.  
With the advent of the GF100-based Fermi GPUs, the maximum number of registers per thread was significantly reduced (i.e., from 255 to 63). 
While the code still worked, the code's efficiency depended on a large number of registers.  Registers
spilled into GPU-device memory and overflowed the L1 cache.
Initially, the performance of Swarm-NG v0.1 was reduced 
when using Fermi GPUs. 
This led us to implement highly parallelized work models. There are four different work models currently used in Swarm-NG:

\begin{itemize}
\item{\Spt: One thread performs all computations 
related to one system.}
\item{\Bpt: One thread is responsible for calculations 
pertaining to one body.}
\item{\Cpt: One thread is responsible for calculations 
related to one coordinate component (i.e., x, y or z) of one body.   
Operations are carried out on the same component of multiple vectors 
(e.g., position, velocity and acceleration).}
\item{\Bppt: One thread is responsible for calculations 
related to one pair of bodies within one \nbody system 
(e.g., mutual interaction between bodies).}
\end{itemize}

While the entire integration could be parallelized using the \spt\  
work model (as in Swarm-NG v0.1), 
the GPU efficiency can be significantly increased using finer-grained 
parallelism.  
Since threads within a thread block can communicate efficiently
using shared memory, we can distribute the calculations for one system
over multiple threads.  
The vast majority of calculations can be parallelized using the
\bpt\ work model ($\nbod$ threads per system).  
While this scheme provides a significant improvement over \spt, 
we found that the performance could be increased further by using 
still finer-grained parallelism.  
For example, for most integrators, the majority of the code for the 
actual integration (i.e., updating the positions and velocities once 
derivatives have been computed) can be parallelized using the 
\cpt\ work model ($3\times\nbod$ threads per system).
One notable exception is the drift step of the MVS integrator that 
uses the \bpt\ work model.  

The most computationally intensive portion of the integration is calculating 
mutual forces between each pair of bodies (specifically the reciprocal
square root) in order to calculate the acceleration of each body.  
For sufficiently small $\nbod$, the mutual component of the force between 
each body pair can be stored in shared memory, so that each force is only
computed once.  
In this case, the \bppt\ work model provides the finest
grain parallelization ($\frac{\nbod(\nbod-1)}{2}$ threads per system) for
calculating the distances between body pairs.  
The final step of accumulating the accelerations acting on a given body 
is to revert to the \cpt\ work model.   

Since no single work model is optimal for all parts of the required 
calculations, Swarm-NG utilizes different work models for different parts 
of the simulation to maximize the GPU efficiency.
The kernel is launched with the maximum number of threads
to accommodate any of the work models. 
We use simple conditional C structures (if statements operating on the 
thread ID) to assign tasks to the appropriate worker threads and 
leave the extra threads idle.  
We issue a \texttt{syncthreads()} when we switch between work models to ensure correct flow of data. Figure \ref{fig:Flow} is an example of this scheme being applied to fixed time step Hermite algorithm with default gravitation implementation for a system with two planets.

We dramatically improved performance of Swarm-NG on Fermi GPUs by implementing the more fine-grained parallelization.  When using the maximum number of threads per multiprocessor, the GF100-based GPUs have fewer register per thread than the GT200-based GPUs.   The new work models is better suited for using fewer registers per thread and still results in higher performance than the original implementation when using GT200-based cards (for most ensemble parameters).  
Further, Swarm-NG's performance is significantly improved using either GT200 or GF100-based GPUs for ensembles with fewer systems, systems with more bodies or more memory intensive integrators.  
We are optimistic that our current implementation will scale well on the next generation of \nv\ GPUs (i.e., codename ``Kepler'').

\subsection{Gravitational force calculations}\label{Sec:Forces}
For CPU-based \nbody integrators, 
the most computationally expensive part of the integration is the force
calculation, as it requires computing a square root to determine
the distance between every pair of bodies.  
Therefore, special attention was paid to optimizing the force calculation
in Swarm-NG.

For small \nbody systems, Swarm-NG uses the \bppt\ 
work model to achieve maximum efficiency when calculating the 
acceleration per unit mass between each pair of bodies exactly once
and in parallel.

For the Hermite integrators, we also calculate the jerk per unit mass
between each pair of bodies.  
The results are stored in shared memory and threads within the block 
synchronize.
Then, the work model reverts to \cpt,
and each thread accumulates the one component of the accelerations
 (and jerks for Hermite integrators) from each of the other bodies.
While the function \texttt{Gravitation::sum()} has
four branches inside a loop, 
the compiler is likely to unroll the loop and optimize the branches.  
Further, all threads within a warp take the same branch of
the remaining \texttt{if} statement.
In this implementation each block requires storing 
$3 \nspb \times \frac{\nbod(\nbod-1)}{2} $ double precision values in the multiprocessor's
 shared memory (twice that amount for the Hermite integrators). 
 Therefore, this version of the force calculation algorithm imposes 
an upper limit on \nbod.  
For $\nbod$ approaching this limit, the efficiency is reduced
since each multiprocessor is limited in the number of blocks that
 it can work with to hide memory latency.  

Swarm-NG also includes an alternative Gravitation class optimized 
for larger $\nbod$.  
This version parallelizes under the \cpt\ work model.
Its force computation calculates each square root twice:
once when calculating the acceleration of body i due to body j
 and once for calculating the acceleration of body j due to body i.  
As a result, each block stores the magnitude of the acceleration 
(and jerk for Hermite integrators) for only $\nbod \times \nspb$ body-pairs into 
shared memory at a time.
Since only $\nbod \times \nspb $ (or $2\nbod\times\nbs $ for Hermite)
double precision values are stored in shared memory, this algorithm
can handle significantly larger $\nbod$.  

In both variants of the force calculation, the impact of the massive star 
(body 0) on acceleration (or jerk) is accumulated last.

\subsection{Integrators}\label{Sec:Integrators}
\begin{algorithm}[h]
\caption{Pseudo code for the generic algorithm highlighting the components
of the framework, including integrator, monitor, force calculation
time step adaptation.}
\begin{algorithmic}
	\STATE $\mathbf{M} \leftarrow$ instance of monitor component for the system
	\STATE $\mathbf{G} \leftarrow$ instance of gravitation component for the system
	\STATE $\mathbf{P} \leftarrow$ instance of propagator component for the system with $\mathbf{G}$ as gravitation component
	\STATE prepare($\mathbf{P}$)
\WHILE{$state_{system} =  $ active \AND $count_{iterations}<max_{iterations}$}
\STATE advance($\mathbf{P}$)
\STATE execute($\mathbf{M}$)
\IF{$time_{system} > time_{destination}$}
\STATE $state_{system} \leftarrow $ inactive
\ENDIF
\STATE $count_{iterations} \leftarrow count_{iterations}  + 1 $
\ENDWHILE
\STATE clean up($\mathbf{P}$)
\end{algorithmic}
\label{Algo:Integrator}
\end{algorithm}

\begin{algorithm}[h]
	\caption{Advance procedure for Hermite propagator. 
	Note that $x^{(0)}_{b,c}$ is the initial $c$th coordinate of $b$th body. Likewise,
	$v^{(0)}_{b,c}$ is the initial $c$th component of velocity of $b$th body.}
	\begin{algorithmic}
		\STATE $ x^{(1)}_{b,c} \leftarrow  x^{(0)}_{b,c} + v^{(0)}_{b,c} \Delta t + \frac{a^{(0)}_{b,c}}{2} (\Delta t)^2 + \frac{j^{(0)}_{b,c}}{6} (\Delta t)^3 $ 
		\STATE $ v^{(1)}_{b,c} \leftarrow  v^{(0)}_{b,c} + a^{(0)}_{b,c} \Delta t + \frac{j^{(0)}_{b,c}}{2} (\Delta t)^2 $

		\FOR{ $k=1$ \TO $2$ }
		\STATE $ a^{(k)}_{b,c}, j^{(k)}_{b,c}  \leftarrow $ calculate acceleration and jerk($\mathbf{G}$)
			\STATE $x^{(k+1)}_{b,c} \leftarrow x^{(k)}_{b,c} - \frac{3}{10} \left( a^{(0)}_{b,c}-a^{(k)}_{b,c} \right) (\Delta t)^2 - \frac{1}{60} \left( 7 j^{(0)}_{b,c} + 2 j^{(k)}_{b,c} \right) (\Delta t) $
			\STATE $v^{(k+1)}_{b,c} \leftarrow v^{(0)}_{b,c} - \frac{1}{2} \left( a^{(0)}_{b,c} - a^{(k)}_{b,c} \right) \Delta t - \frac{1}{12} \left( 5 j^{(0)}_{b,c} +  j^{(k)}_{b,c} \right) (\Delta t)^2 $
		\ENDFOR

		\STATE $ time'_{system} \leftarrow time_{system} +  \Delta t $
	\end{algorithmic}
\label{Algo:Hermite}
\end{algorithm}

Swarm-NG includes several integrators that build on the common data structures and work models described above.  
This framework makes it easy for a developer to implement a new integrator that takes advantage of the highly parallel GPU architecture with minimal attention to hardware details.  
First, we describe the common pattern for Swarm-NG integrators.
Then, we discuss implementation details relevant to specific integrators.

Swarm-NG provides a generic GPU integrator class that facilitates rapid development and testing of various integration algorithms and/or parallelization schemes.  
The generic integrator class provides common logic, such as setting which threads are active and which are idle for various portions of the calculation, generating references and pointers to the system, logging the system state, and enforcing the stopping criteria.
A simplified pseudocode is listed in Algorithm \ref{Algo:Integrator}.
A ``propagator'' class is responsible for advancing the system by one time step
(e.g., Hermite propagator in Algorithm \ref{Algo:Hermite}).
A ``monitor'' class is responsible for determining when the system state should be logged and when the GPU should cease integrating a given system.  
The generic integrator automatically enforces a maximum number of iterations of the propagator and a maximum destination time, separate from the monitor class.

After some initial setup, the generic GPU integrator kernel initializes the monitor and propagator while providing monitor and propagator their respective parameters.
Next, it enters a loop allowing for many propagator steps within a single GPU kernel call, so as to minimize overhead associated with each kernel call. 
For each iteration, the generic integrator calculates the maximum time step, calls the propagator's \texttt{advance()} function, synchronizes threads,  calls the monitor function and resynchronizes the threads.  
If the monitor determines that logging is needed, then the system's current state is written to a buffer in the GPU global device memory.  
The system can be set inactive either by the monitor class or if the system reaches the destination time.  

While the generic GPU integrator class allows for developers to focus on their algorithm with minimal attention to bookkeeping and GPU hardware, it incurs some additional overhead, and imposes some constraints on the structure.  
The overhead can be significant for relatively simple integrators and few-body systems (e.g., Hermite).  
Therefore, Swarm-NG also offers the developer the option of writing a GPU integrator kernel that bypasses the generic GPU integrator class.  
The developer is still able to avoid many details of GPU programming by inheriting from a GPU integrator base class and reusing data structures and programming patterns from existing integrators.

\subsubsection{Hermite Integrator}\label{Sec:HermiteImpl}
The Hermite integrator closely follows the pattern of the generic GPU integrator.
There are three differences.
A thread loads the position and velocity of the body-component that it is assigned once prior to entering the loop over multiple integration steps, rather than loading the data at the beginning of each integration step.  
Since Hermite has a relatively small integration kernel, removing the additional loads improves performance.
Second, the Hermite integrator has been optimized to remove \texttt{syncthreads()} calls that are needed in the generic GPU integrator to ensure correctness, but are not necessary for the Hermite integrator.  
With the reduced number of synchronization statements, the GPU has more flexibility in assigning threads to hide memory latency.

Finally, the Hermite integrator requires both the acceleration and the jerk be computed analytically.  
Therefore, it must use a gravitation class that computes both the acceleration and jerk.  
As a result, the Hermite integrator uses more shared memory per block than the Verlet, Runge-Kutta and MVS integrators, so fewer systems can be assigned to a multiprocessor simultaneously.  
As a result either the number of systems per block or the maximum number of blocks executing on a multiprocessor is reduced relative to an integrator which uses a Gravitation class that does not require computing the jerk.

\paragraph{Time Step adaptation}
For adaptive time steps, the user specifies 
a constant step size scale ($\tau$).  
During each step, the simulation time is advanced by a time step 
($\Delta t$) that is recalculated based on the ratio of the acceleration
and the jerk for each body,
\be
\Delta t = \tau \left[ \sum_i \frac{\left|{\bf j}_{i,0}\right|^2}{\left|{\bf a}_{i,0}\right|^2}\right]^{-\frac{1}{2}},
\ee
where $i$ indexes a body in the system.
Each acceleration and jerk component for each body is computed
by the \texttt{Gravitation} class.  
These values are written to shared memory and the ratio
$\left|{\bf j}_{i,0}\right|^2/\left|{\bf a}_{i,0}\right|^2$ 
is determined using the body per thread work model.  
The summation is performed using the system per thread work model
and the result is distributed to all threads using shared memory.

Since the above time step algorithm uses only the values at the beginning
of the time step, technically, it is not fully time-symmetric.  
In principle, one could ensure time symmetry by iterating until
the time steps calculated using the coordinates at the beginning
and end of the step are the same.  
To facilitate convergence, one could choose from a discrete set of time steps,
e.g., $\tau \times 2^q$ for any integer $q$ \cite{Kokubo1998}.  
Such a time step scheme could be valuable if the Hermite integrator
were to be used for long-term integrations.  
However, complete time symmetry is often not essential for short-term
integrations, the most common use case for the Hermite integrator.

\subsubsection{Runge-Kutta Integrator}\label{Sec:RKCKImpl}
The fixed time step Runge-Kutta integrator implementation is very similar to that of the fixed time step Hermite integrator.  
Since Runge-Kutta uses the acceleration (and not the jerk), it uses less shared memory per system than the Hermite integrator, potentially allowing a multiprocessor to integrate more systems simultaneously.
However, the Runge-Kutta integrator requires more than three times as many computations per step (i.e., six rather than three force calculations per step).  
Furthermore, the Runge-Kutta integrator requires much more memory for local variables than the Hermite integrator.  
The local variables may not fit in registers, and for practical block sizes the local variables may not even fit into the L1 cache on current Fermi GPUs.
Therefore, local variables spill over into the L2 cache and perhaps global device memory.  
This results in significantly greater latency in accessing memory and reduces computation throughput.  

\paragraph{Time Step Adaptation}
The adaptive version of Runge-Kutta integrator uses the Cash-Karp
method at the end of the step to compute the 5th order error. If the
error exceeds the user spec- ified accuracy parameter \texttt{error
tolerance} ($\epsilon$), The step is rejected, e.g, positions,
velocities and time are not updated in the device memory. Instead, the
next attempted time-step will use a smaller time-step.

\subsubsection{MVS Integrator}

The Mixed-Variable Symplectic (MVS) integrator is implemented as a propagator to be used with the generic GPU integrator.  
The four sub-steps that use Cartesian coordinates (both Cartesian drift sub-steps and both kick sub-steps; see \S\ref{Sec:MvsTheory}) use the body-component per thread work model.
The force calculation is implemented the same as for Runge-Kutta and uses the body-pair per thread work model for small \nbod.
The Keplerian drift sub-step (drifting along a Keplerian orbit; see \S\ref{Sec:MvsTheory}) and the variable transformations are performed using the body per thread work model.
Since the separation between planets does not change during Cartesian drift and kick sub-steps, we reuse the accelerations calculated after Keplerian drift substep at the beginning of the next step.


%
%
%
%
%
%
%
%
%

\subsection{Optimizations} \label{Sec:Optimisations}
\def\chunk{{\em{chunk}}}

\paragraph{Coalesced Memory Access}

In current CUDA architecture, the threads within a block are assembled into ``warps'', a group of $\nws$ threads ($\nws = 32$ on current \nv GPUs) that are executed (nearly\footnote{For some operations, e.g. trigonometric or sqrt functions, a Special Function Unit is used.  Since there are more threads than SFUs, threads must be queued for execution on SFUs.}) concurrently.  

Although only one warp can be executed at once, multiprocessor can schedule warps to execute in a manner that hides memory latency for load and store operations.
Load/store operations can be significantly faster if threads within a warp access data in a contiguous region of memory.  
This is known as ``coalesced'' memory access.   

Swarm-NG provides data structures ({\texttt{CoalescedStructArray}} and {\texttt{CoalescedMemberArray}}) that facilitate coalesced memory access and generally optimize the memory access pattern.  
Using these classes, accessing these arrays correctly is transparent to developers. 
In the data structures, systems are grouped into ``chunks''.  
The data for the same member of a structure for all systems within a chunk is stored consecutively, so as to facilitate coalesced memory access.
The number of systems in a chunk ($\ncs$) must be power of 2 and can be set at compile time to a number between 1 and 32. 
For fully coalesced memory access, $\ncs$ should be chosen in a way that
$\ncs \ge \nws$ and $\nspb = m \times \ncs$ for some integer $m$.

The non-trivial data structure minimizes the cost of load/store operations by allowing for both coalesced memory access and good cache performance (as compared to a structure of arrays indexed by the system number).

\paragraph{Coherent Execution}

The chunk-based data structures also have implications for the organization of threads within a block.
If multiple threads within a warp take different branches (e.g., an \texttt{if} statement or loop), then the time required is as if all threads executed every branch taken by any of the threads within the warp.  
Therefore, maximum efficiency is achieved when a warp is ``coherent'', meaning that all threads within the warp execute the same instructions by taking the same branch of conditional statements or loops.  

Within a block of threads, threads are organized into a two dimensional thread grid, where the dimensions of the outer array is the number of threads per system ($\nthread$) and the dimension of the inner dimension is the number of systems per block.  
This implies that systems within a chunk that access the same structure member  are accessing a continuous region of memory, allowing them to benefit from coalesced memory access.  
Additionally, the threads are organized so that underutilized threads are ordered consecutively, allowing the GPU to exit from the extra threads efficiently.  
This occurs whenever the current work model uses fewer than the total number of threads launched for the kernel (e.g., during the calculation of distances for $\nbod \le 6$).

\paragraph{Caching in Fermi architecture}
A second rule of thumb for optimizing GPU code is to maximize memory throughput.
In general, transfer to and from global memory is expensive due to the high memory latency ($\sim$300-1000 clock cycles if not cached).  
The shared memory and L1 cache have much smaller latencies (e.g., $\sim$20 clock cycles).  
Swarm-NG uses data structures optimized to provide for both coalesced memory access and higher spatial and temporal locality.
This provides significantly better performance on an architecture where cache is present (e.g., Fermi GF100 GPUs) with minimal cost for older GPUs (e.g., GT200) that do not cache GPU memory.

\paragraph{GPU-optimized mathematical functions}
We take advantage of the GPU hardware-optimized \texttt{rsqrt}(x) (reciprocal square root) function which is slightly faster than 1.0/\texttt{sqrt}(x). Although \texttt{rsqrt}(x) is not IEEE-754 complaint the deviation does not introduce significant numerical error. 
Similarly, the MVS integrator makes use of the combined \texttt{sincos}(x) function.

\paragraph{Aggressive loop unrolling using C++ templates}
We unroll small loops (e.g., over bodies or body pairs) so as to avoid loop overhead and to provide the compiler information at compile time that allows 
further optimization of code flow.  
For some loops, we use the \texttt{\#pragma} compiler directive to unroll loops.
However, the compiler may choose to ignore this directive.  
On the other hand, the C++ compiler is obliged to unroll templates, guaranteeing that sequential code will be generated for loops and code flow can be optimized at compile time.  
Therefore, we use template meta-programming in order to generate static, optimized code for the most important loops, e.g., the loop over body pairs in the gravitation class.  
The static code generation gives a better performance at the cost of longer compile time and a significantly larger code size.

\paragraph{Data Reduction}
For some applications, transferring data from the GPU memory to CPU memory can become a bottleneck.
Data reduction is the strategy of choosing to download only a subset of data and/or performing post-processing on the GPU, so less data needs to be transferred to CPU, reducing CPU-GPU communication overhead. 

When using the simplest syntax, Swarm-NG synchronizes the state of the ensemble between the CPU and GPU before and after each kernel call.  
Advanced programmers may choose to call the \nbody integrator kernel without invoking CPU-GPU memory transfer, provided they manage the CPU-GPU memory transfer separately.
This can be useful for applications that frequently inspect the state of the systems, but do not need all data about every system.  
For example, when comparing model planetary systems to radial velocity observations, one only need to inspect the star's radial velocity, rather than every component of every body.  Transferring only these coordinates reduces the amount of memory transferred by more than a factor of six.  
Another example consists of monitoring the energy and/or angular momentum of each system.
The energy and angular momentum can be calculated on the GPU, so only 1-4 values per system need to be transferred to the CPU.

\paragraph{Automatic selection of $\nspb$ and register utilization}
Our CUDA kernels utilize the maximum number of registers allowed per thread
in Fermi architecture (63).  Each multiprocessor has a finite size
register file, so register utilization has considerable
effect on overall performance of the simulation. 
At run-time, Swarm-NG automatically selects the block size to maximize $\nspb$~ 
considering the following three hardware resource limits:
1) the maximum number of threads allowed per multiprocessor,
2) the size of shared memory module per multiprocessor, and
3) the size of the register file per multiprocessor. 
The automatic selection is largely based
on our performance benchmarks explained in \S\ref{Sec:Performance}.
For a given chunk size, Swarm-NG's performance generally increases with increasing occupancy, which is defined to be the ratio of active warps to the maximum number of warps supported on a multiprocessor.  For GT200-based GPUs, Swarm-NG achieves near 25\% occupancy for standard integrators.  For Fermi-based GPUs, Swarm-NG achieves nearly 30\% occupancy for the standard integrators. 
If one wants to choose the optimal values for $\ncs$ and $\nspb$, then
we recommend performing benchmarks specifically for the problem of interest. 
For example, in some cases, we observe a modest performance benefit when using a smaller $\nspb$, if that choice also results in an occupancy comparable to or greater than that of the maximum $\nspb$.  
This can be due to slightly greater occupancy (due to requirement that $\nspb$~ is a multiple of the chunk size) or due to reducing the impact of synchronization operations.

\section{Validation and Performance}
\label{Sec:Results}
In this section, we present the results of \nbody simulations
that were designed to 
(a) verify that the GPU integrators in Swarm-NG are implemented 
correctly and to 
(b) measure how the performance of the integrators depends
on simulation and implementation parameters.
\S\ref{Sec:InitialConditions} lists the initial conditions for our 
tests and benchmarks,
\S\ref{Sec:Validation} explains how integrators are validated.
\S\ref{Sec:Performance} presents performance benchmarks.

\subsection{Initial Conditions for Verification \& Benchmarking}
\label{Sec:InitialConditions}
The range of possible initial conditions and the
chaotic nature of the general \nbody problem
make it impossible to test the integrators in all possible regimes. 
Since the accuracy and the limitations of the underlying integration 
algorithms are well-studied \cite{Kokubo1998,Press:1992:NRC:148286,Wisdom1991},
Swarm-NG uses the regime of particular scientific interest to the authors:
the integration of planetary systems.  

Here, we consider a system with one central body, the `sun',
with mass $m_0=1$ placed at rest at the origin and
the \npl `planets' indexed by $i=1$ with identical masses $m_i$ and
initial conditions corresponding to a circular orbit about
 the origin with separation of $a_i :=(1+\alpha)^{i-1}$.  
Thus, the initial velocity is perpendicular to the position vector and
has a magnitude $|v_i| = 2\pi(\frac{m_0}{a_i^3})^{\frac{1}{2}}$.
Unless specified otherwise, our benchmarks are based on the case $\nbod=3$.
Default values are relative planet masses comparable to Jupiter,
 $m_i=0.001$ (\texttt{planet\_mass} parameter in the configuration file) 
with $\alpha=0.4$ (\texttt{spacing\_factor} parameter).
For $\npl=2$, these initial conditions are provably `Hill stable',
i.e., there will be no close approaches, and hence provide a test
appropriate for both fixed and adaptive time step algorithms -- 
and they are reasonable for $\npl>2$,
where long-term orbital stability can not be guaranteed,
even if equations of motion were integrated exactly.

\subsection{Validation}\label{Sec:Validation}
\subsubsection{Direct Comparison of Results}
To compare two integrators using the same initial conditions,
the Swarm-NG demonstrate code offers a `test' mode,
where a user can provide a set of initial conditions and 
a set of reference final conditions.  
This test mode allows users to test existing integrators in new regimes 
or to validate the new integrators.
Swarm-NG performs the integration and compares the magnitude of the 
Cartesian position and velocities of each body (relative to the origin), 
as well as the final integration time.  
Passing the test requires that all match to within specified tolerances.  
The default value for position tolerance (\texttt{pos\_threshold}),
velocity tolerance (\texttt{vel\_threshold}) and time tolerance
(\texttt{time\_threshold}) is $10^{-10}$.

First, we verified agreement of short-term simulations using 
a CPU-based Hermite integrator and a GPU-based Hermite integrator.
Differences below the thresholds are due to hardware precision and/or 
rounding behavior such as our GPU implementation using the 
hardware-optimized \texttt{rsqrt} (reciprocal square-root) function rather 
than \texttt{sqrt}.
Next, to verify the accuracy of the Swarm-NG integrators,  
rather than implementing a CPU version of each integrator,  
we compared their results to that of the GPU-based Hermite integrator.  
When comparing the results of two different integration algorithms, 
we use a sufficiently small time step or accuracy parameter
so that the truncation error is minimized for both integrators.
Since the actual systems are chaotic, we must restrict comparisons
to short-term integrations.  
The Swarm-NG demonstration code's default test integration duration is $10\pi$ time units (or 5 inner orbital periods) and can be overridden using the \texttt{destination\_time} configuration parameter.  
All of Swarm-NG's integrators pass these tests using the initial conditions described in \S\ref{Sec:InitialConditions} and the default integrator parameters for $\nbod=3\ldots 6$.

\subsubsection{Energy Conservation}
\label{Sec:EConservation}

To test the long-term stability of Swarm-NG GPU integrators,
we monitored the total energy of a closed system evolving
under Newtonian gravity. For an exact integrator,
the system energy is preserved. The energy is given by
\be
E=\sum_i m_i \frac{1}{2} \left|{\bf v}_i\right|^2 - \sum_{i\ne j} 
\frac{G m_i m_j}{\left|{\bf r}_i-\bf{r}_j\right|},
\ee
where $\bf{r}_i$ and $\bf{v}_i$ are the position and velocity vectors
of the $i$th planet.
We monitor $\left|\frac{\Delta E(t)}{E(0)}\right| = \left|\frac{E(t)-E(0)}{E(0)}\right|$,
the fractional error in the energy of an integration as a function
of the time in the simulation.

\begin{figure}[h]
  \includegraphics{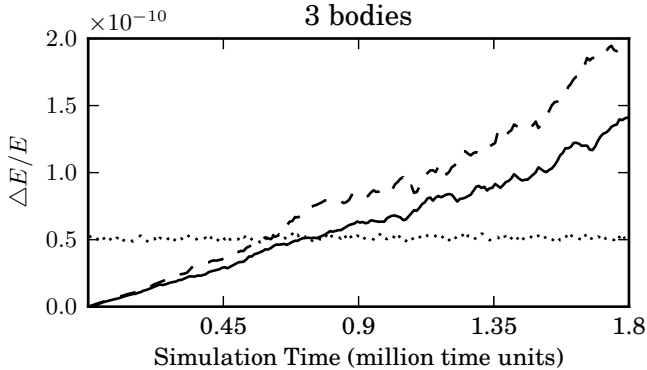}
  \caption{Energy error versus time.  
  The three line styles correspond to simulations using the adaptive time step Hermite (solid), adaptive time step Runge-Kutta (short, but thick dashes) and MVS (dotted).  
  Initial conditions are described in \S\ref{Sec:InitialConditions}.  
  The time step and accuracy parameters are listed in \S\ref{Sec:EConservation} and have been chosen so that the integrators have a roughly similar level of energy error after $\sim 100,000$ inner orbital periods.    
}
  \label{fig:Conservation}
\end{figure}

We ran simulations for $2 \times 10^{6}$ time units, 
corresponding to over 286,000 inner orbital periods.  
After experimenting with the time-step and accuracy parameters, we adopted 
$\Delta t = 10^{-3}$ for Fixed time-step Hermite, 
$\Delta \tau = 1.7 \times 10^{-2}$ for Adaptive time-step Hermite, 
$\epsilon = 5.0 \times 10^{-33}$ for Adaptive time-step Runge-Kutta, and
$\Delta t = 10^{-2}$ for MVS.
We use these values for all future benchmarks.  
These choices result in a very good long-term energy conservation and a roughly similar level of energy error after $\sim 100,000$ inner orbital periods, facilitating rough comparisons of the performance of different integrators over these timescales.

\begin{table}[h]
	\begin{center}
		\begin{tabular}{lcccc}
			& \multicolumn{4}{c}{Simulation Time (million time units)}   \\ \cline{2-5}
			Integrator & .45 & .90 & 1.3 & 1.8 \\ \hline
			Hermite Adaptive & 2.9 & 6.3 & 9.2 & 14.1 \\
			MVS & 5.0 & 5.2 & 5.1 & 5.1 \\
			Runge-Kutta Adaptive & 3.7 & 8.6 & 12.2 & 19.2 
		\end{tabular}
	\end{center}
	\caption{Fractional energy error for selected integration algorithms. 
Results ($\times 10^{-11}$) are listed for $\nbod=3$ using initial conditions as described
in \S\ref{Sec:InitialConditions}.  The columns indicate the energy error after different integration times.  Time-step parameters are listed
in \S\ref{Sec:EConservation}.}
	\label{table:Conservation}
\end{table}

The growth of the fractional energy error for selected integrators is shown in Figure \ref{fig:Conservation} and specific values are given in Table \ref{table:Conservation}.

As expected, the Hermite and 
the Runge-Kutta integrators result in a slow accumulation of energy error,
while the MVS integrator results in a roughly constant energy error.  
We verified that the Swarm-NG integrators are capable of high-precision
integrations; most scientific studies use significantly larger time-steps 
and lower accuracy.

\subsection{Performance}\label{Sec:Performance}

\begin{table*}
	\begin{center}
	\begin{tabular}{lcccc}
		Machine           & Cluster 1     & Cluster 2      & Server 1      & Workstation    \\ \hline
		CUDA Device       & Tesla T10     & Tesla M2070    & Tesla C2070   & Geforce GTX480 \\
		CUDA Driver       & 4.2           & 4.2            & 4.2           & 4.2            \\
		CUDA Toolkit      & 4.2           & 4.2            & 4.2           & 4.2            \\
		Compute Capability& 1.3           & 2.0            & 2.0           & 2.0            \\
		GPU MPs           & 30            & 14             & 14            & 15             \\ 
		Cores/MP          & 8             & 32             & 32            & 32             \\
		GPU Clock         & 1.30 GHz 	  & 1.15 GHz       & 1.15 GHz      & 1.45 GHz       \\
		GPU Memory        & 4GB           & 6GB            & 5GB           & 1.5GB          \\
		Memory Clock      & 800 MHz       & 1566 MHz       & 1494 MHz      & 1900 MHz       \\
		Shared Memory     & 16KB          & 48KB           & 48KB          & 48KB           \\
		L1 Cache Size     & None          & 16KB           &  16KB         &  16KB          \\
		L2 Cache Size     & None          & 768KB          & 768KB         & 768KB          \\
		Registers per MP  & 16384         & 32768          & 32768         & 32768          
	\end{tabular}
	\end{center}
	\caption{Key specifications of GPUs used for performance benchmarks.   }
	\label{Tab:GpuSpecs}
\end{table*}

The performance of GPU integrators in Swarm-NG depends 
on the number of systems, $\nsys$,  the number $\nbod$ 
of bodies in each system, as well as on the parameters \nbs, \ncs\ and
of course on the GPU hardware.  
In this section, we examine how the performance of selected GPU integrators 
is affected by these parameters.  
The results can help users optimize their choice of implementation parameters, 
or at least avoid particularly inefficient choices.  
The results also demonstrate that several of the optimizations implemented
in Swarm-NG provide significant performance benefit compared to a naive GPU
implementation. 

Unless noted otherwise, performance benchmarks are based on integrating an ensemble of 
2000 planetary systems for 1 time unit, $\nbod=3$, $\ncs=4$ and an automatically 
chosen number of systems per block (\nspb); see \S\ref{Sec:Optimisations}.

\begin{figure}[htb]
     \includegraphics{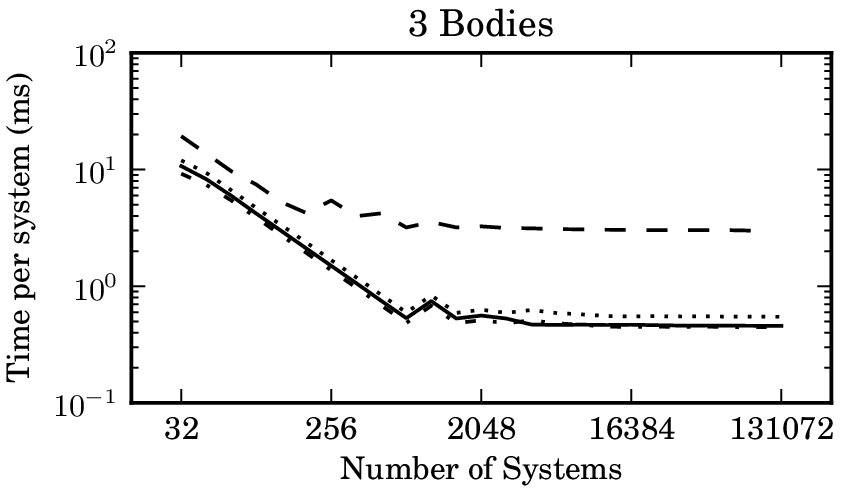}
     \includegraphics{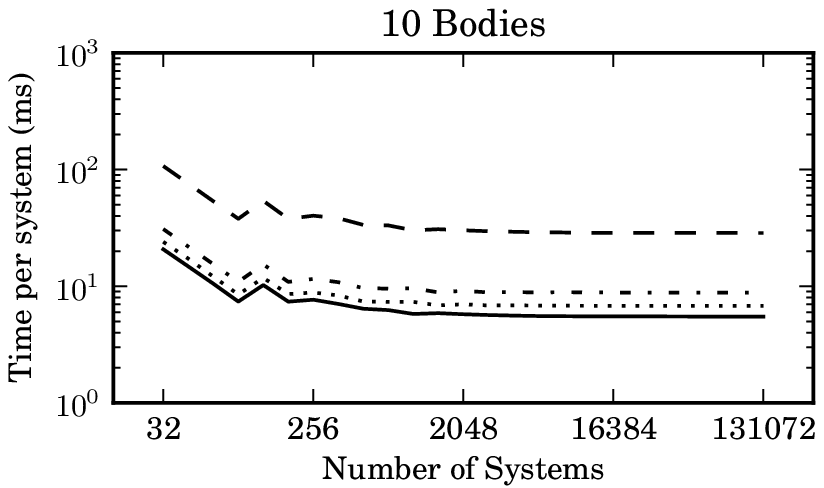}
     \caption{Wall clock time per system versus number of systems in
     the ensemble for the Hermite (fixed time step) integrator and
     the standard initial conditions and integration parameters 
     described in \S\ref{Sec:InitialConditions} and \ref{Sec:EConservation}.
    The line styles show results for different machines: 
    Cluster 1 (dashed), Cluster 2 (dotted), Server 1 (dot-dashed),
    Workstation (solid).}
		\label{fig:Comparison}
\end{figure}

\begin{table}[h]
	\begin{center}
	\begin{tabular}{lrrrr}
		     &     \multicolumn{4}{c}{Machine}                     \\ \cline{2-5}
		$n$  & Cluster 1  & Cluster 2  & Server 1    & Workstation \\ \hline
		3    &  140.008   & 35.951     & 29.290      & 30.482      \\ 
		4    &  203.036   & 65.284     & 52.635      & 53.132      \\ 
		5    &  310.685   & 100.800    & 69.346      & 81.880      \\ 
		6    &  807.346   & 121.908    & 102.880     & 103.536     \\
		10   & 1874.480   & 444.202    & 577.674     & 362.020
	\end{tabular}
	\end{center}
	\caption{Performance comparison of different Machines. The numbers in
	table represent the time (in seconds) required to integrate 
	the ensemble.
	The ensemble in this benchmark consists of 131072 systems and 
	it was integrated for 1 time unit 
	using the default configurations of the Hermite integrator.}
	\label{Tab:GpuComparison}

\end{table}

Fig.\ \ref{fig:Comparison} and Table \ref{Tab:GpuComparison} compare 
the performance of the Hermite and MVS integrators using three different 
types of GPU hardware  whose specifications are given in 
Table \ref{Tab:GpuSpecs}.

Figure \ref{fig:Comparison} shows that
the wall clock time for integrating each system asymptotes 
for large \nsys, as expected since both computation and memory access
scale linearly.
However, when there are not enough systems in the ensemble
to utilize all GPU resources, the time per system 
significantly increases.  On the other hand, the wall clock time per system is minimized when the ensemble contains enough systems to efficiently utilize the GPU's computational resources while waiting for memory transfers.  
The minimum number of systems to achieve near maximum efficiency 
usually scales with the number of multiprocessors.
On the `Server 1' configuration, high efficiency is reached when
the ensemble size is $\nsys>1024$ for a 3-body system,
respectively $\nsys>128$ for a 10-body system.

\begin{figure}[htb]
  \includegraphics{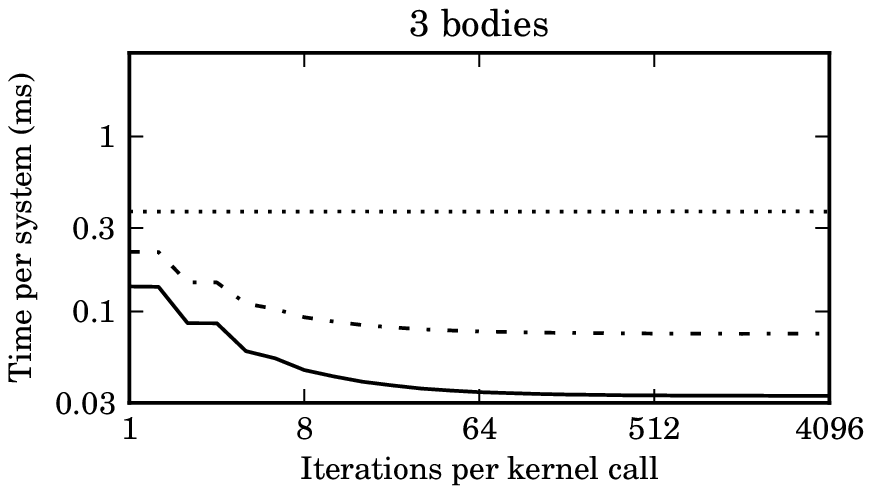}
  \includegraphics{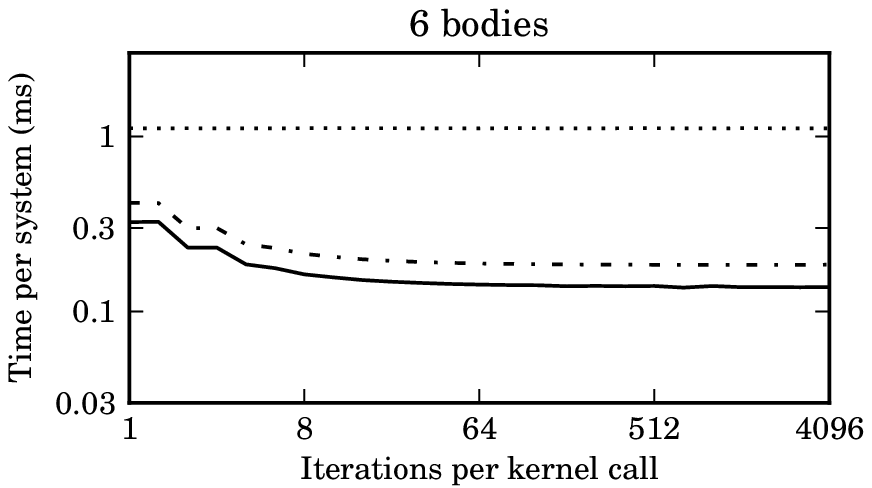}
  \caption{Wall clock time per system versus number of integrator iterations per GPU kernel call.  
  The line styles show results for three different integrators, Hermite (solid), MVS (dotted), Runge Kutta (dotted-short dashed).  
  Integrations were performed on Server 1 using standard initial conditions and integration parameters described in \S\ref{Sec:InitialConditions} \& \ref{Sec:EConservation}.
	}
  \label{fig:MaxIterations}
\end{figure}
Next, we test our design choice of carrying out all of the 
computations and flow control on the GPU device.
Unlike most GPU accelerated applications,
Swarm-NG performs the entire \nbody integration on the GPU.  
The host program needs only set up parameters, 
load initial configuration onto the GPU,
call the GPU-based integrator, and examine the results.  
Once control is transferred to GPU, the simulation runs for a 
configurable number of iterations, i.e., steps for Hermite or MVS integrators
or trial steps for Runge-Kutta integrator. 
 
Figure \ref{fig:MaxIterations} shows, for Hermite and MVS integrators,
that the wall clock time per system 
decreases as the number of iterations of the \nbody integrator increases.
As expected, there are significant costs associated 
with each kernel call 
as demonstrated in the figure; the speed-up is $\sim~2.4\ldots 4$ if each kernel call performs hundreds of steps.
For the Runge-Kutta integrator, the run-time is insensitive to the number 
of iterations. Therefore, it could have been implemented as efficiently using
a separate kernel call for each step of the integrator.

While part of this overhead is the kernel launch itself and some operations
are performed once per thread (e.g., calculating memory location for the system being 
integrated by a given thread), we posit that most of the performance
decrease is due to the memory latency,
i.e., the GPU not being utilized while waiting for the initial conditions to be
loaded from the device global memory.  
Presumably, the Hermite and MVS integrators benefit from multiple steps per 
kernel call, since they are less memory intensive, and hence
initial conditions remain in the cache between successive kernel iterations.  
More iterations per kernel call lead to longer kernel calls which hide the
latency.

We conclude that there can be a substantial performance benefit to performing 
the integration fully on the GPU.

\begin{figure}[h]
 \includegraphics{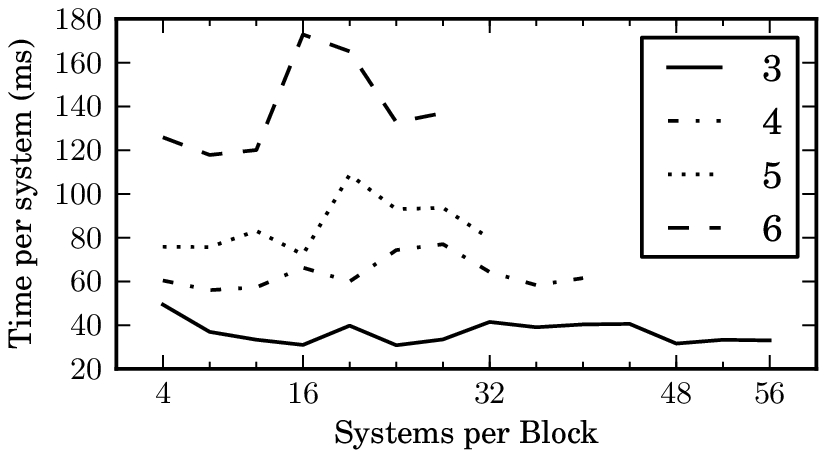}
  \caption{Wall clock time per system versus number of systems per block.
  The line styles show results for different number of bodies per system, 
  as indicated in the legend.
  The maximum number of systems per block varies with the number of bodies
  per system, due to the available shared memory and the maximum number of 
  threads per block.  
  Integrations were performed on Server 1 using standard initial conditions
  and integration parameters described 
  in \S\ref{Sec:InitialConditions} and \ref{Sec:EConservation}.
  }
  \label{fig:BlockSize}
\end{figure}

\paragraph{Coalesced Memory Access}
Figure \ref{fig:BlockSize} shows wall clock time per system varying
with $\nspb$.   
The number of systems per block must be an integer multiple of the 
chunk size, $\ncs$.
Here we set $\ncs=4$, allowing for at least partially coalesced memory access and several different values of $\nspb$.  
For $\nbod=3$, the performance increases as $\nspb$ approaches 16, 
likely to due the benefits of fully coalesced memory access 
and utilization of each warp.
On the other hand, for $\nbod=6$, the performance decreases once $\nbs=16$, 
corresponding to the local memory within a block exceeding 16kB
and potentially resulting in cache misses or limiting the number of
threads that can be active at once in a multiprocessor.  
Thus, the variations in the time per system with $\nspb$ is likely due to
a combination of factors: memory caching and coalescence, and 
efficient utilization of threads.

\begin{figure}[h]
  \includegraphics{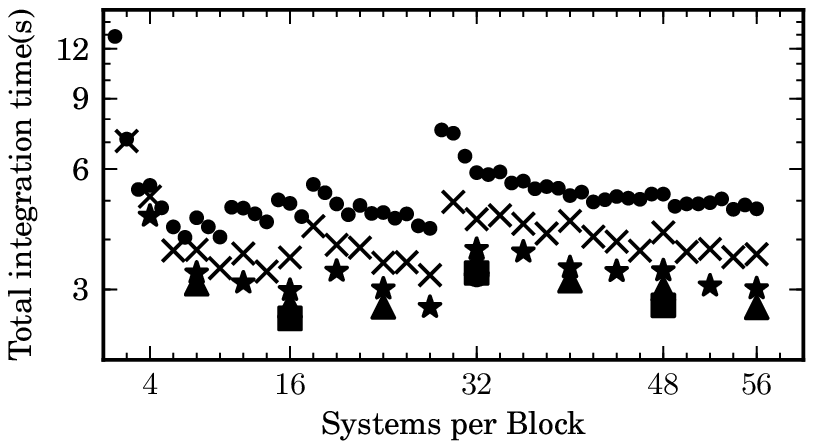}
  \caption{
  Wall clock time per system versus number of systems per block for \nbod=3 and the fixed time-step Hermite integrator.   
  The point styles show results for different values of the chunk size, 16 (squares), 8 (triangles), 4 (stars), 2 (\texttt{x}'s) and 1 (disks).  
  Integrations were performed on Server 1 using standard initial conditions and integration parameters described in \S\ref{Sec:InitialConditions} 
  and \ref{Sec:EConservation}.
  }
  \label{fig:ChunkSize-BlockSize}
\end{figure}

Figures \ref{fig:ChunkSize-BlockSize} shows how the performance of the Hermite integrator with $\nbod=3$ depends on both $\nspb$ and $\ncs$.  
The most significant performance benefits (from $\nspb={1\dots 8}$) come from grouping systems into chunks which results in better memory coalescing. 
There is no significant gain when increasing $\ncs$ from 16 to 32, indicating that a chunk size of 16 exploits the maximum amount of memory coalescing.

Since the GPU is configured to allocate 64 registers per thread,
the maximum number of threads that can be launched on a multiprocessor
is $ \frac{32768}{64} = 512$. 
Since the total number of threads per block 
is 252 for $\nspb=28$ and 504 for $\nspb=56$, 
these choices of $\nspb$ result in optimal register utilization and
hence best performance.

\newpage
\begin{figure}
  \includegraphics{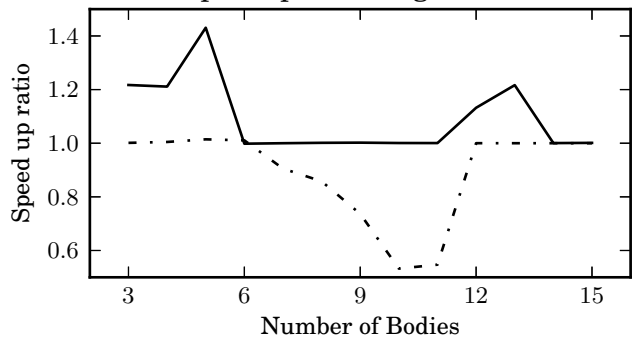}
  \caption{Ratio of wall clock time for integrations 
  when the CUDA driver is instructed to favor 
  (a) a configuration with 48kB L1 cache and 16kB shared memory per
  multiprocessor  and
  (b) a configuration with 16kB L1 cache and 48kB shared memory per 
  multiprocessor.  
  Ratios greater than one indicate higher performance of configuration (a).
  Solid line refers to Server 1 and dotted-dashed line refers to Cluster 2.
  Integrations used the standard initial conditions and integration parameters
  described in \S\ref{Sec:InitialConditions} and \ref{Sec:EConservation}}
  \label{fig:CacheSize}
\end{figure}

Fermi-based GPUs have 64kB of memory per multiprocessor that can be partitioned
among shared memory and an L1 cache in two ways.  This memory shares a common 
hardware implementation, but is accessed differently. 
The shared memory (either 16kB or 48kB) is user-managed, i.e., it is only 
accessed as instructed by the programmer. 
Swarm-NG uses shared memory for calculating accelerations or managing
adaptive time steps, depending on the integrator and choice of gravitation 
class (see \S\ref{Sec:Implementation}). 
While the programmer has no control over how the L1 cache is used, 
it can help hide the latency of local memory access automatically. 
Figure \ref{fig:CacheSize} compares Swarm-NG's performance 
for the default configuration but with varying balance of shared memory to
L1 cache.

\begin{figure}[h]
  \includegraphics{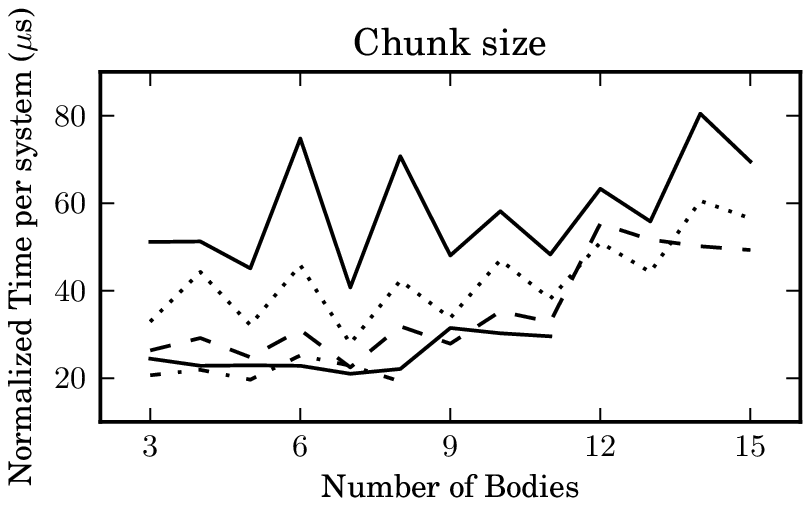}
  \includegraphics{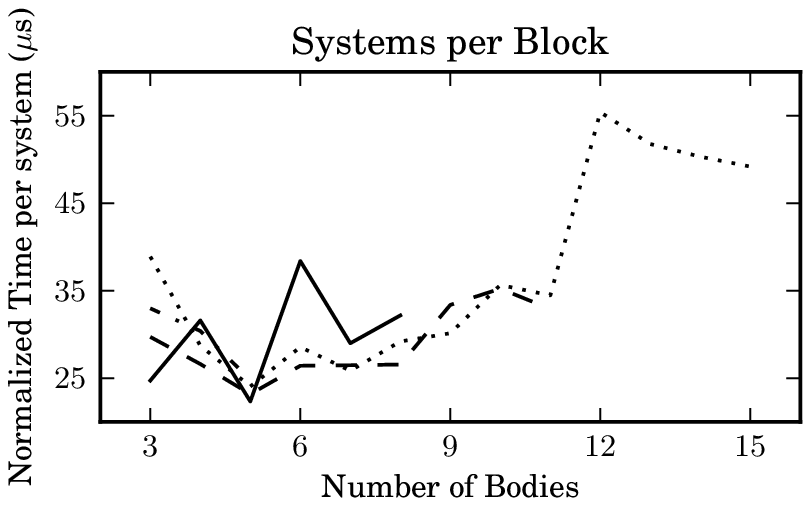}
  \caption{Normalized wall clock time per system versus number of bodies per system using the adaptive time step Hermite integrator.   
  Here ``normalized'' refers to dividing the wall clock time per system by the number of body pairs.  
  In the top panel, the line styles show results for different values of \ncs, 1 (upper solid), 2 (dotted), 4 (dashed), 8 (lower solid) and 16 (dot-dashed).
  In the bottom panel, the line styles show results for different values of \nspb, 4 (dotted), 8 (dashed), 16 (solid) and 32 (dot-dashed).
  Integrations were performed on Server 1 using standard initial conditions and integration parameters described in \S\ref{Sec:InitialConditions} \& \ref{Sec:EConservation}.}
   \label{fig:NBody_GpuParam}
\end{figure}

Figure \ref{fig:NBody_GpuParam} shows the wall clock time per system of the Hermite integrator normalized by the number of body pairs versus the number of massive bodies in each system. By default, values for $\ncs$ and $\nspb$ are chosen automatically. However, one can
run benchmarks similar to ones in Figures 
\ref{fig:ChunkSize-BlockSize} 
and 
\ref{fig:NBody_GpuParam} 
to find the optimal parameters.  We find that for fixed $\nbod$ and $\ncs$, the occupancy is usually an accurate predictor of the relative performance when using two different values of $\nspb$.

\subsubsection{Comparison to OpenMP implementation on multi-core CPU}

\begin{table}[h]
\begin{center}
\begin{tabular}{ccc}
	NBodies & CUDA   & OpenMP \\
	\hline 
	3 & 102.1	 & 45.8 \\
	4 & 115.5	 & 53.5 \\
	5 & 128.8	 & 58.5 \\
	6 & 156.8	 & 61.7
\end{tabular}
\end{center}
\caption{Numbers of interactions per second (in millions) for different implementations of the Hermite integrator.  One interaction is the force calculation between a pair of planets.  Benchmarks include time spent on the integration, as well as the force calculation.  }
\label{table:CPUGPUIntPerSec}
\end{table}

\begin{figure}[h]
  \includegraphics{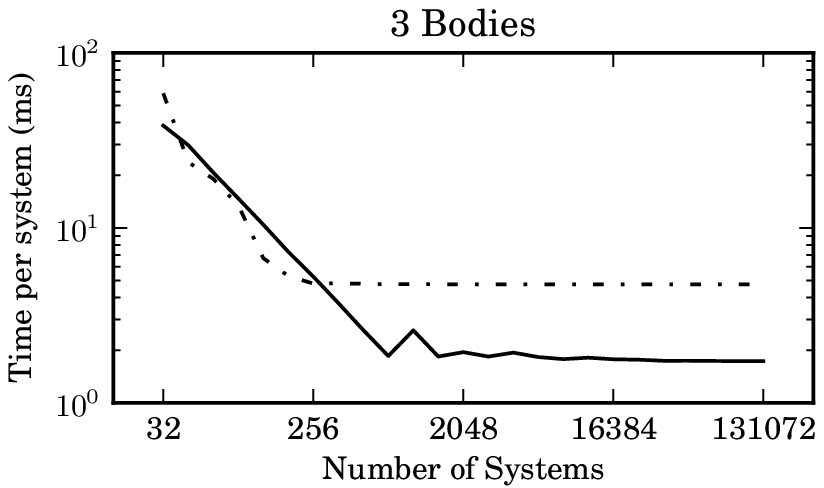}
  \caption{
  Comparison of GPU implementation(solid line) and CPU implementation of Hermite integrator(dotted-dashed).The integration is run for 10 time units with different number of 3-body systems.  
  Integrations were performed on Cluster 2 using standard initial conditions and integration parameters described in \S\ref{Sec:InitialConditions} \& \ref{Sec:EConservation}.}
   \label{fig:CPUGPUComparison}
\end{figure}

In the beginning, we implemented a simple implementation of Hermite integrator in CPU as a reference implementation. The main purpose of the CPU implementation is to verify correctness of GPU implementation. For this reason, we tried to avoid hand optimization and complex features
in CPU implementation to avoid logical errors.

However, to give a fair comparison between GPUs and multi-core CPUs, we enhanced the CPU implementation with OpenMP API to maximize performance on multi-core CPUs. We keep the code simple and rely on Intel optimizing compiler to generate fast code (same approach is used in GPU implementation with CUDA compiler).

OpenMP enhanced implementation is parallelized on a \spt\ basis. The code is compiled with Intel optimizing compiler version 12.1.5 and following flags are used to optimize the code on a 6-core Intel Xeon X5675 CPU: \texttt{-O3 -fast -xCORE-AVX-I -fopenmp}. 

Table \ref{table:CPUGPUIntPerSec} compares performance of the OpenMP implementation versus  the GPU implementation in terms of interactions per second. The benchmark was generated by integrating an ensemble of 131072 systems each containing 3-bodies  for 100000 steps in Hermite integrator.
Figure \ref{fig:CPUGPUComparison} compares the wall clock time of the GPU and CPU-based Hermite integrators for 3-body systems.

\section{Discussion}
\label{Sec:Discussion}
Swarm-NG enables fast parallel simulation of
thousands of few-body systems using modern GPUs. 
The Swarm-NG package includes a high-level API
and powerful demonstration programs, 
already in use for science. 
Code samples illustrate Swarm-NG's modular nature 
and how to easily add functionality such as custom data logging,
stopping conditions, user-specified force laws and 
even custom integrators.  

Swarm-NG permits hand-optimizing modules that try to boost 
performance of every aspect (i.e., integration, force calculation, data logging, stopping criteria) for specific applications.   We recommend these be used sparingly
for two reasons.  First, such optimization means
writing software specific to the hardware whose architecture changes
from time to time. In such case, the code need to be revised (or rewritten) for optimized 
performance on the new architecture -- a cumbersome process as
we experienced when migrating from GT200 to GF100.
Second, modularity of the code is important to allow non-CUDA-savvy
users to choose from a variety of integration schemes, force laws, stopping
conditions and logging options.  Indeed,
one of the major achievements of Swarm-NG is to leverage
compiler optimization and template meta-programming to achieve a highly optimized code,
rather than compromising the clear high-level structure of the code. 

\subsection{Comparison to Odeint}
The only alternative project we are aware of,
that offers scientific grade integrations of ODEs on GPU, is  
Odeint\footnote{{http://headmyshoulder.github.com/odeint-v2/index.html}}.
Odeint is a C++ framework for solving ODEs based on template
meta-programming and can integrate systems using either the CPU or
the GPU \cite{DBLP:journals/corr/abs-1110-3397}. 
Support for integrating ODEs on the GPU is provided via
Thrust\footnote{{https://code.google.com/p/thrust/}}, a  template library for 
very high-level parallel programming on GPUs and multi-core CPUs 
(see Ch. 26 of \cite{Hwu:2011:GCG:1964878}).
Early in the development, we envisioned a design similar to Odeint.
However, we found that such generality came at a significant performance loss.  
For example, Swarm-NG's performance has been 
significantly improved by careful layout of data.
While code parallelized by Thrust can often results in coalesced memory access,
improving cache performance through data locality
(see \S\ref{Sec:Optimisations}) with Thrust would be quite difficult
to achieve. 
As another example, Thrust does not allow users to take advantage 
of shared memory.  
For systems with small $\nbod$, the use of shared memory reduces the number 
of square root calculations in the force calculation step by a factor of 6, 
compared to Odeint with Thrust.  
So, while very-high-level frameworks like Odeint and Thrust are excellent
for rapid code development, common applications such as \nbody
integration can benefit greatly from being implemented directly in CUDA.

\subsection{Applications}
\label{Sec:Applications}
Few-body integration is a ubiquitous tool 
in  computational astrophysics and planetary science.  
Swarm-NG has already proven useful for several applications related to the evolution of planetary systems.  

One class of applications is related to studying planet formation
in a general sense, such as characterizing the stability of
 tightly-packed planetary systems (B\'{e}dorf et al.\ in prep.) 
or investigating the effects of stellar flybys on planetary systems 
\cite{2012arXiv1204.5187B}. 
Swarm-NG makes it practical to integrate a large number of 
hypothetical planetary systems, required to characterize how the orbital 
evolution depends on the initial conditions.
Another class of applications involves studying actual planetary systems. 
For example, by assuming that billion year-old planetary systems are 
long-term stable, Swarm-NG has placed limits on the masses and 
eccentricities of planetary systems discovered by NASA's Kepler mission
\cite{2012arXiv1206.4718C,2012Natur.482..195F}.
A sort of hybrid scenario involves integrating a large ensemble of
 planetary systems similar to an actual planetary system, but varying one 
or two of the initial conditions to determine how close the actual system 
is to an unstable region of phase space.  NASA's Kepler mission recently
 discovered the first planetary systems containing a planet orbiting not one,
 but a pair of main-sequence stars.  Swarm-NG was used to show that a modest
 reduction in the semi-major axis of the planet would render these systems 
unstable \cite{2012arXiv1206.4718C,2012Natur.481..475W}. 

One key application of Swarm-NG is for the self-consistent analysis 
of astronomical observations of extrasolar planetary systems. 
 Most confirmed extrasolar planets have been discovered by measuring a pattern
 of changes in the velocity of the host star via Doppler spectroscopy. 
 A Bayesian framework provides the basis for inferring physical parameters 
(and their uncertainties) from astronomical observations.  Markov chain Monte 
Carlo (MCMC) methods are now routinely used for rigorous analyses of single 
planet systems \cite{2005AJ....129.1706F}.  The analysis of systems 
with multiple planets is much more computationally demanding,
 both because of the higher-dimensional parameter space and the need 
to perform an \nbody integration for each model evaluation.  
For some multiple-planet systems, the orbital motion can be well-approximated
 as the linear superposition of the motion of each star-planet pair 
\cite{2006ApJ...642..505F}.  However, for other systems, mutual planetary 
interactions produce significant observable consequences even on observable
 timescales, so direct \nbody integrations are necessary to properly 
interpret the observations 
\cite{2011AJ....141...16J,2001ApJ...551L.109L,2006ApJ...641.1178L,2001ApJ...556..296M,2011ApJ...729...98P,2011ApJ...730...93W}.  
While standard MCMC algorithms are serial in nature, population-based MCMC
 algorithms can take advantage of GPUs ability to integrate many planetary 
systems in parallel.  
In particular, the Differential Evolution Markov chain Monte Carlo (DEMCMC)
 algorithm \cite{ter2006markov} is well suited to GPUs 
and often more efficient than a standard MCMC algorithm even when executed on a
 single CPU.
E.B.F. has developed a DEMCMC code for the self-consistent analysis of 
Doppler observations of multiple planet systems.  This code has already 
been applied to several planetary systems
\cite{2011AJ....141...16J} (Wang et al., submitted to ApJ; 
Lee et al., in prep.; Nelson et al., in prep.). 
 The \nbody integrations can be performed using either OpenMP or a GPU, 
using the Swarm-NG library.  In the case of the 55 Cnc planetary system with
 5 planets, the DEMCMC simulations required roughly three weeks, even using
 a Tesla C2070 GPU.  Thus, the self-consistent Bayesian analysis of this system
 was simply not practical prior to Swarm-NG.  

The other common method for discovering extrasolar planets involves
 measuring the apparent brightness of the host star decrease when a planet
 passes in front of the star.  As for systems discovered by the Doppler method,
 Bayesian analysis of a single planet system is relatively straight forward
 using either a standard MCMC \cite{2010exop.book...55W} or DEMCMC 
algorithm \cite{2012arXiv1206.5798E}.  Recently, NASA's Kepler mission has 
discovered hundreds of systems with multiple transiting planets.  Indeed,
 some of the most interesting systems contain multiple closely-spaced planets 
that demand direct \nbody integration to model properly. 
 An analysis package combining direct \nbody integration and 
the DEMCMC algorithm was implemented using MPI and a cluster of CPUs. 
 While this code has been applied to several systems
 \cite{2012arXiv1206.4718C,2011Sci...331..562C,2012Natur.481..475W}, 
even a two planet system can require tens of thousands of CPU hours. 
 As the Kepler mission is continuing to return data and uncover additional 
planets, the computational requirements for self-consistent Bayesian analysis 
of Kepler data will become even more challenging.  Therefore, we plan to
 develop a GPU-based code for performing DEMCMC analyses of Kepler 
observations, including direct \nbody integrations performed on the GPU 
via Swarm-NG.  

While Swarm-NG was developed with a focus on planetary systems, 
it can be readily applied to variety of other problems, such as a system of moons 
orbiting a planet, scattering of multiple star systems, or even a swarm 
of stars orbiting a black hole.


\section*{Acknowledgements}
We thank Alice Quillen, as well as Mark Harris and David Luebke of \nv\ Corp. for helpful discussions about implementing and optimizing CUDA code. 
We thank Craig Warner and Ying Zhang for helping to improve the organization and documentation of the Swarm-NG code.
We thank Jeroen B\'{e}dorf,  Sourav Chatterjee, Constanze R\"{o}dig and Mariusz Slonina for helping to test early versions of Swarm-NG.  
E.B.F. acknowledges the Lorentz Center for their hospitality and facilitating collaborations that helped improve Swarm-NG.
This research was supported by NASA Applied Information Systems Research Program Grant NNX09AM41G.  
The authors acknowledge the University of Florida Research Foundation's Research Opportunity Seed Fund for supporting the early stages of this research.
The authors acknowledge the University of Florida High-Performance Computing Center for providing computational resources and support that have contributed to the research results reported within this paper.

\bibliographystyle{model2-names}
\bibliography{collection}

\end{document}